# Peer Effects and Herd Behavior: An Empirical Study Based on the "Double 11" Shopping Festival

Hambur Wang

Guanghua School of Management, Peking University, Beijing 100871, China

**Abstract**: This study employs a Bayesian Probit model to empirically analyze peer effects and herd behavior among consumers during the "Double 11" shopping festival, using data collected through a questionnaire survey. The results demonstrate that peer effects significantly influence consumer decision-making, with the probability of participation in the shopping event increasing notably when roommates are involved. Additionally, factors such as gender, online shopping experience, and fashion consciousness significantly impact consumers' herd behavior. This research not only enhances the understanding of online shopping behavior among college students but also provides empirical evidence for e-commerce platforms to formulate targeted marketing strategies. Finally, the study discusses the fragility of online consumption activities, the need for adjustments in corporate marketing strategies, and the importance of promoting a healthy online culture.

## 1. Introduction

In recent years, despite a slowdown in China's overall economic growth, the rapid development of the internet economy has injected new vitality into the sustained and stable growth of the nation's economy. With continuous advancements in internet technologies and the rise of e-commerce platforms, online shopping has become an indispensable part of people's daily lives. The "Double 11" shopping festival, a unique large-scale online shopping event in China, takes place every year on November 11th, attracting a large number of consumers and becoming a globally recognized economic phenomenon.

The complexity of consumer behavior is particularly evident during the "Double 11" shopping festival. Purchasing decisions are influenced not only by factors such as price, product quality, and promotional activities but also by peer effects and social trends. Especially during such large promotional events, peer effects and herd behavior have a significant impact on consumers' purchasing decisions and overall market behavior. Understanding these influencing mechanisms is crucial for both academic research and practical applications.

Existing literature on consumer behavior predominantly focuses on aspects such as price, product quality, and promotional strategies, with relatively limited research on peer effects and herd behavior (He Jia, 2013). However, with the widespread use of social media and online communities, consumers are increasingly inclined to rely on the opinions and behaviors of their peers when making decisions (Xi Mingming, Zhu Limeng, 2016). Therefore, studying peer effects and herd behavior is of significant theoretical and practical importance in understanding consumer decision-making processes in the online shopping environment.

This paper employs Bayesian estimation of peer effects and uses data from 200 valid questionnaires to investigate the herd behavior and its characteristics during the "Double 11" shopping festival from the perspective of offline peer effects. The Bayesian Probit model, by incorporating prior and posterior distributions, effectively addresses potential endogeneity issues in the model, making it suitable for robust estimation under small sample conditions and providing uncertainty information for parameter estimates.

## 2. Analytical Method

### (1) Theoretical Basis and Methodological Considerations for Choosing the Bayesian Probit Model

The Bayesian Probit model, by incorporating prior and posterior distributions, effectively addresses potential endogeneity issues that often arise in economic and social science research. One of its significant advantages is its ability to integrate prior information, providing a more flexible and precise model specification. In the context of a sample size of 200, the Bayesian approach remains robust in small sample conditions, ensuring reliable estimation results. Moreover, this method offers uncertainty information through posterior distributions, allowing researchers to gain a more comprehensive understanding of the credibility of the estimation results.

The Bayesian Probit model is particularly suitable for handling binary dependent variables, making it an excellent tool for capturing the nonlinear effects of peer effects and herd behavior (Yi Xiaoyun, 2019). By incorporating prior information and utilizing posterior analysis, this model not only enhances the reliability of estimation outcomes but also strengthens the explanatory power for complex behavioral patterns. Therefore, the Bayesian Probit model has significant advantages and broad applicability in the study of peer effects and herd behavior.

## (2) Specific Relationship Formulas

### 1. Consumer Decision-Making Process

First, it is necessary to describe the consumer decision-making process in detail to clarify how consumers make purchase decisions based on personal information and observations of previous consumer behavior. This process involves a sequential decision-making model: the first consumer makes decisions entirely based on personal information, while subsequent consumers reference the behavior of previous consumers during their decision-making (Xing Ziyan et al., 2019). By applying the Bayesian Probit model, we analyze in depth the psychological logic of the third consumer, who is inclined to maintain decision consistency after observing the identical decisions of the first two consumers. In this sequential decision-making model, each consumer's decision is not only based on their own information but is also significantly influenced by the behavior of previous decision-makers. This phenomenon is referred to as "information cascade" in economics and behavioral sciences, which suggests that individuals often rely on others' decisions as a reference to reduce decision-making risks when facing uncertainty. The Bayesian Probit model, by incorporating prior and posterior distributions, effectively captures this complex decision-making behavior pattern and addresses potential endogeneity issues. The specific relationship formula is as follows:

$$p\left(i^* = i^{1,2} \mid \text{roommate}\right) = \frac{\text{p}\left(\text{roommate} \mid i^* = i^{1,2}\right) \cdot p\left(i^* = i^{1,2}\right)}{\int \text{p}\left(\text{roommate} \mid i^* = i^{1,2}\right) p(i)}$$

Where $i^*$ represents the final decision of the consumer, and *roommate* indicates the influence of peer effects. This formula, through the Bayesian probability updating mechanism, captures how consumers adjust their decisions under the influence of peer effects.

Through this model specification, we can not only quantify the influence of peer effects on consumer decisions but also reveal the dynamic changes in herd behavior under different scenarios. The estimated results of the model will provide important empirical evidence for understanding the complexity of consumer behavior during the "Double 11" shopping festival and offer theoretical support for e-commerce platforms to devise more targeted marketing strategies.

**2. Bayesian Equilibrium**

In a scenario where the first two consumers have made identical decisions (either participating or not participating in the "Double 11" shopping event), if the third consumer's information diverges from the choices of the first two, according to the Bayesian Probit model, the third consumer is more likely to follow the decisions of the first two consumers rather than stick to their original intent. This is because, in this situation, the probability of choosing to follow the strategy with correct information, based on Bayesian equilibrium, exceeds the probability of sticking with one's initial choice (Xi Mingming, Wu Zhijun, 2020):

$$\pi^*_{i_{1,2}|\text{ roommate}} = \alpha^3\beta^2(1-\beta) + \alpha^2\beta(1-\beta)(1-\alpha)/p(\text{ roommate}) >$$
$$\pi^*_{i_3|\text{ roommate}} = \alpha^2\beta(1-\beta)(1-\alpha)\beta/p(\text{ roommate})$$

In this model, α and β represent the consumer's trust levels in personal information and peer behavior, respectively. This equilibrium state reflects how consumers balance personal information and peer behavior to make optimal decisions under conditions of information asymmetry. The Bayesian Probit model, through its probability updating mechanism, details the decision adjustment process of consumers under peer effects, revealing the significant influence of peer behavior on consumer choices.

Through this analysis, we gain a deeper understanding of consumer decision-making logic in online promotional events, particularly under the influence of peer effects, providing important evidence for further theoretical research and practical applications. This result has significant reference value for e-commerce platforms' marketing strategies and for government policies related to consumer protection.

## 3. Data Sources and Measurement

### (1) Variable Explanation

In this study, we selected a series of independent variables that may influence "Double 11" online shopping behavior to comprehensively analyze consumer behavior patterns. These variables include peer effects, online shopping experience, gender, fashion consciousness, dormitory relationships, household registration, poverty status, major, grade, and income.

First, peer effects, specifically the consistency of roommate behavior, are used to measure the extent to which consumers are influenced by the behavior of their roommates. Second, online shopping experience is a key variable that examines whether consumers have engaged in online shopping in the past. Gender is used to explore potential differences in consumer behavior. The fashion consciousness variable reflects consumers' attitudes toward considering "Double 11" shopping as a fashionable activity, while the dormitory relationship variable investigates the impact of the harmony level within dormitory relationships on consumer decisions. Household registration (urban or rural) is another important variable for analyzing its potential impact on consumer behavior. The poverty status variable examines whether the consumer is classified as a low-income student, to understand how their economic status influences purchasing decisions. The major variable focuses on differences between students majoring in economics and management and those in other disciplines, investigating its impact on consumer decisions. The grade variable studies how consumer behavior changes at different academic stages. Finally, the income variable analyzes how consumers' monthly income or allowance levels influence their purchasing decisions.

Through a comprehensive analysis of these variables, we can gain deeper insights into the factors affecting consumer decisions, providing theoretical support for e-commerce platforms' marketing strategies and government consumer policies. This multi-dimensional analysis approach not only helps reveal the complexity of "Double 11" shopping behaviors but also offers new perspectives for further research in related fields.

## (2) Descriptive Statistics

To investigate consumers' herd behavior and its characteristics during the "Double 11" shopping event, we conducted a questionnaire survey and collected 200 valid responses, yielding a response rate of 93.46%. The survey was primarily conducted online, with questionnaires distributed through WeChat groups to university students at several universities in Beijing. To encourage participation, each respondent had the opportunity to randomly receive a red envelope, with amounts ranging from 0.5 to 1 yuan. The specific data are shown in Table 1.

## Table 1: Variable Selection and Variable Explanations

| Variable Name | Variable Definition | Mean | Maximum | Minimum |
|---|---|---|---|---|
| Participation | Whether the respondent participated in the "Double 11" shopping event: 1 = participated, 0 = did not participate | 0.834 | 0 | 1 |
| Peer Effects | Whether the roommate's behavior is consistent with the respondent: 1 = consistent, 0 = inconsistent | 0.811 | 0 | 1 |
| Online Shopping Experience | Whether the respondent has online shopping experience: 1 = has experience, 0 = no experience | 0.984 | 0 | 1 |
| Gender | Gender of the respondent: 1 = female, 0 = male | 0.592 | 0 | 1 |
| Fashion Consciousness | Whether the respondent considers "Double 11" shopping a fashionable activity: 1 = agrees, 0 = disagrees | 0.266 | 0 | 1 |
| Dormitory Relationship | Whether the dormitory relationship is harmonious: 1 = harmonious, 0 = not harmonious | 0.812 | 0 | 1 |
| Household Registration | Household registration: 1 = urban, 0 = rural | 0.637 | 0 | 1 |
| Poverty Status | Whether the respondent is a registered low-income student: 1 = registered, 0 = not registered or not low-income | 0.239 | 0 | 1 |
| Major | Major: 1 = economics or management, 0 = other majors | 0.631 | 0 | 1 |
| Grade | Year of study: 1 = freshman, 2 = sophomore, 3 = junior, 4 = senior, 5 = graduate student | 1.843 | 0 | 5 |
| Income | Monthly income or allowance: 1 = ≤1000 yuan, 2 = 1001-2000 yuan, 3 = 2001-3000 yuan, 4 = 3001-5000 yuan, 5 = >5000 yuan | 2.493 | 1 | 5 |
| Returns | Whether the respondent has had return experience: 1 = yes, 0 = no | 0.487 | 0 | 1 |

## 4. Results and Analysis

### (1) Empirical Test of Herd Behavior from the Perspective of Peer Effects

Based on the Bayesian Probit model and its inference results discussed earlier, this study conducts an empirical test of herd behavior from the perspective of peer effects. To facilitate analysis, both the traditional Probit model and the Bayesian Probit model are used to estimate herd behavior in "Double 11" online shopping consumers. The table below presents the empirical test results from the perspective of peer effects.

检 Table 2: Empirical Test of Herd Behavior from the Perspective of Peer Effects

| Variable | Traditional Probit | | | Bayesian Probit | | |
|---|---|---|---|---|---|---|
| | （1） | （2） | （3） | （4） | （5） | （6） |
| Peer Effects | 0.262*** | 0.178*** | 0.186*** | 0.235*** | 0.194*** | 0.221*** |
| | （0.0298） | （0.0278） | （0.0311） | （0.0249） | （0.0308） | （0.0144） |
| Online Shopping Experience | | 0.391*** | 0.357*** | | 0.392*** | 0.381*** |
| | | （0.0427） | （0.0357） | | （0.0438） | （0.0287） |
| Gender | | 0.0845*** | 0.0902*** | 0.0798*** | | 0.085*** |
| | | （0.0144） | （0.0129） | （0.0178） | | （0.0162） |
| Fashion Consciousness | | 0.408** | 0.338*** | 0.342*** | | 0.326*** |
| | | （0.0285） | （0.0238） | （0.0302） | | （0.0175） |
| Control Variables | YES | YES | YES | YES | YES | YES |
| Regional Effects | YES | YES | YES | YES | YES | YES |
| Constant Term | YES | YES | YES | YES | YES | YES |
| Observations | 204 | 204 | 204 | 204 | 204 | 204 |

**Note:** Values in parentheses are standard errors. *Significance levels:* ***p < 0.01, *p < 0.05*.

According to the estimation results from both the traditional Probit model and the Bayesian Probit model, the marginal effect of peer effects (roommate) ranges from 18.6% to 23.5%, and all results are significant at the 1% level. This indicates that the participation of a roommate significantly influences the participation decision of other consumers during the "Double 11" shopping event. Specifically, when consumers face peer effects, the probability of participating in the "Double 11" shopping event increases by 18.6% to 23.5% compared to consumers whose roommates did not participate.

This conclusion highlights the significant positive influence of peer effects on online shopping behavior. Based on these results, it can be inferred that during the "Double 11" shopping festival, the participation of roommates significantly increases the likelihood of other consumers participating in the event. This phenomenon demonstrates the profound impact of peer effects on consumer herd behavior, especially in large promotional events, where consumers are more likely to adjust their decisions based on the behavior of those around them.

## (2) Analysis of Calculation Results

### 1. Distribution of Participation in the Event

This section presents the distribution of respondents who participated in or did not participate in the "Double 11" online shopping event. According to the data, 83.4% of the respondents participated in the "Double 11" shopping event, while the remaining 16.6% did not. This high participation rate suggests that "Double 11" has become a widely embraced online shopping activity, with its marketing strategies and promotional events effectively attracting a large number of consumers.

Specifically, the proportion of respondents who participated in "Double 11" is significantly higher than that of non-participants. This difference in participation rates underscores the importance of "Double 11" in the minds of consumers and its immense market influence. Through effective marketing and compelling promotional strategies, the "Double 11" shopping festival has successfully stimulated consumer purchasing desire, demonstrating its dominant position in the field of e-commerce.

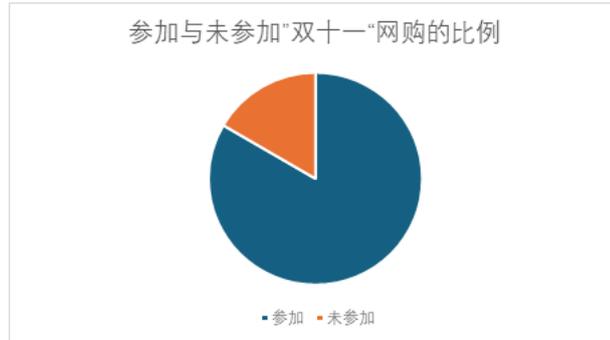

## 2. The Impact of Peer Effects on Participation in Activities

The chart compares the proportion of respondents who participated in the "Double 11" (Singles' Day) online shopping event under conditions of peer effects (i.e., when roommates' behaviors align) versus no peer effects. The data reveals that 81.10% of respondents participated in online shopping when peer effects were present, compared to only 18.90% when peer effects were absent. The significantly higher participation rate of 81.1% compared to 18.9% suggests that peer effects play a facilitating role in influencing "Double 11" shopping behavior.

This indicates that consumers' decisions are significantly influenced by the behaviors of those around them, demonstrating a tendency for conformity. Specifically, when a consumer's roommate also participates in the "Double 11" shopping event, the likelihood of the consumer participating increases significantly. This behavioral pattern highlights the critical role of peer effects in online shopping and further confirms the widespread presence of the conformity effect in consumer decision-making. The following chart illustrates the proportions of participation in "Double 11" shopping with and without peer effects:

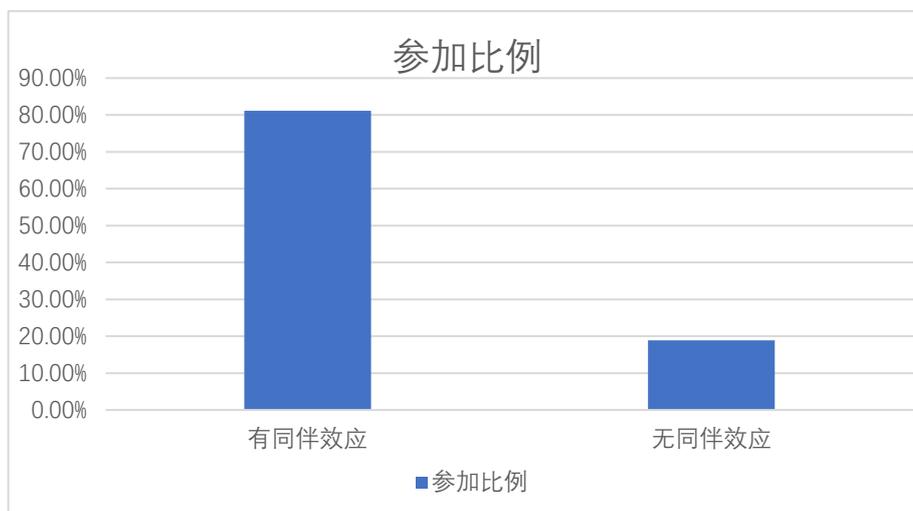

The chart clearly demonstrates the influence of peer effects on consumer participation decisions, further validating the strong impact that peer behavior has on individual decision-making. Future research could explore the manifestation of this effect across different consumer groups, as well as the moderating roles of other potential variables such as age, gender, and income level. A multidimensional analysis of these factors would provide a more comprehensive understanding of the mechanism by which peer effects operate in consumer behavior, offering empirical evidence to support the development of more effective marketing strategies.

**3. The Relationship Between Gender and Participation in Activities**

According to the survey data in this study, female participants represent the majority in the "Double 11" (Singles' Day) online shopping event, accounting for 59.2% of the sample. Compared to men, women exhibit stronger tendencies toward conformity behavior in several aspects.

Firstly, women tend to gather together to communicate, discuss, and share shopping experiences. In such interactions, their thoughts, behaviors, and emotions are more likely to resonate with one another, leading to a consensus and making them more susceptible to peer influence, thus exhibiting conformity behavior. Secondly, women are more inclined to purchase experiential products such as beauty and skincare items, as well as clothing and accessories. These products typically require personal use to assess their effectiveness, making women more vulnerable to the influence of others' evaluations and recommendations, which fosters conformist consumer behavior. Lastly, women place greater importance on the group's recognition of their identity. When exposed to peer effects, women are more likely to compromise in order to meet the expectations and standards of the group. This psychological tendency makes them more prone to conformity behavior in specific contexts.

The following table presents the participation proportions and conformity behavior characteristics of women and men in the "Double 11" online shopping event:

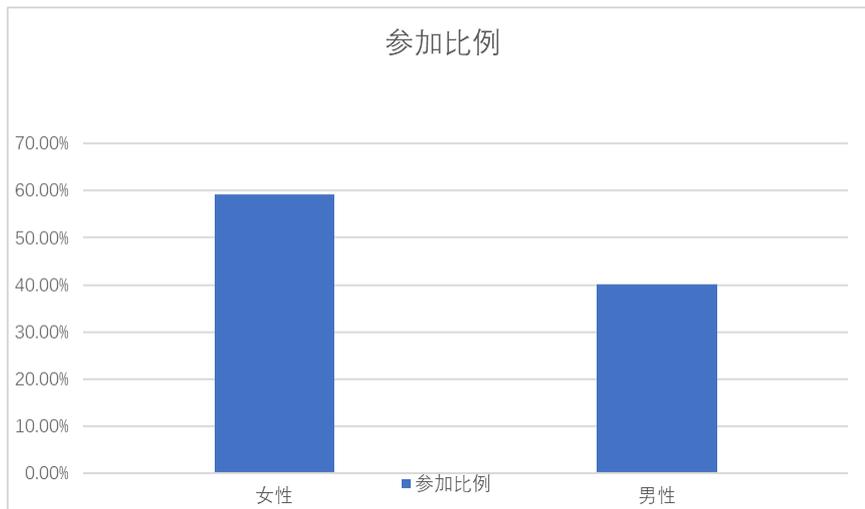

This finding indicates that women exhibit a stronger tendency toward conformity behavior during the "Double 11" online shopping event, which is closely related to their higher levels of social interaction, preference for experiential products, and greater emphasis on group identity. This discovery provides a new perspective for understanding gender differences in online shopping behavior. Future research could explore other factors that may influence this gender gap, such as age, education level, and socioeconomic status.

## 5. Conclusion and Implications

### (1) Conclusion

This study first reveals the significant impact of peer effects on consumer decision-making. The data shows that when a student's roommate participates in the "Double 11" shopping event, the student's likelihood of participating increases by at least 18%. This result remains robust even after rigorous statistical corrections using instrumental variables, firmly demonstrating that peer behavior has a direct and profound impact on university students' consumption choices during shopping events such as "Double 11."

Secondly, although conformity behavior plays an important role in consumer decision-making, analysis indicates that university students do not entirely disregard their independent judgment. Their conformity behavior varies depending on the context. For example, students with less online shopping experience or those who do not view "Double 11" as a fashionable event show a significantly increased likelihood of participating in shopping due to peer influence. This differentiated conformity phenomenon reveals how university students, in their pursuit of maximizing personal utility, rationally integrate peer influence and personal preferences.

Finally, the study also examines the influence of gender and other individual characteristics on conformity behavior. The results show that, under the influence of peer effects, female consumers are on average 10% more likely to exhibit conformity behavior compared to male consumers. Additionally, factors such as major, hometown type, personality traits, and family economic status did not exhibit significant differences in conformity behavior. This finding further emphasizes how consumers balance personal judgment and social influence when making rational conformity choices.

Through these findings, this study not only deepens our understanding of the psychological and behavioral patterns of university students during shopping festivals like "Double 11," but also provides empirical evidence for e-commerce platforms to develop more precise marketing strategies. These results offer valuable theoretical support for e-commerce operational strategies, helping platforms more effectively leverage peer effects to optimize marketing campaigns and increase consumer participation.

## (2) Implications

### 1. The Vulnerability of Online Consumer Decision-Making

This study reveals the high sensitivity of online consumer behavior to negative signals. This sensitivity stems from the nature of rational conformity behavior, distinct from the irreversibility of information cascades. Once consumer satisfaction declines, if spread through influential channels like social media influencers or authoritative organizations, it can rapidly trigger changes in consumer behavior. This phenomenon, referred to as the "tide-out effect," may severely disrupt the online economy, especially in the context of insufficient domestic demand. Therefore, to ensure the healthy and stable development of the online economy, regulatory bodies must enforce strict supervision and penalties for fraudulent advertising, substandard products, and other improper practices.

### 2. **Adjustments in Corporate Marketing Strategies**

When planning online promotional activities, businesses should not solely rely on online reviews and word-of-mouth effects. Instead, they should comprehensively consider the offline peer effects and word-of-mouth impacts. This indicates that while managing online reputation, companies must also focus on product quality and offline marketing activities. Only when online and offline reputations complement each other can companies achieve long-term stable sales growth. Therefore, companies should fully leverage peer effects and word-of-mouth effects in their marketing strategies to enhance overall sales performance.

3. **Promoting a Healthy Online Consumption Culture**

Society should advocate for a healthy online consumption culture through various channels, making it a cultural trend. This not only enhances the consumer shopping experience, allowing them to enjoy discounts while also enjoying the joy of shopping, but also brings positive energy to society, promoting the healthy development of the online economy. As active participants in online consumption, university students should cultivate independent judgment and critical thinking skills. When faced with overwhelming promotions and advertisements, they should learn to distinguish between truth and falsehood, avoid blind conformity, and make rational consumption decisions. At the same time, students should actively engage in the construction of online culture by sharing positive consumer experiences and evaluations to help others make better consumption choices, thus enhancing their sense of social responsibility. In the future, network content recommendation systems based on deep learning methods (Wang et al. 2024 & Zhao et al. 2024 & Yao et al. 2024) such as transformer (Wang et al., 2024) could be introduced to promote a healthy online consumption culture.